# Post-stishovite transition in hydrous aluminous SiO$_2$

Koichiro Umemoto[1,2], Katsuyuki Kawamura,[3] and Kei Hirose[2,4,5], and Renata M. Wentzcovitch[2,6,7]

[1]*Department of Earth Sciences, University of Minnesota, 310 Pillsbury Drive SE, Minneapolis, MN 55455, USA*

[2]*Earth-Life Science Institute, Tokyo Institute of Technology, 2-12-1-IE-12 Ookayama, Meguro-ku, Tokyo 152-8550, Japan*

[3] *Department of Sustainable Resource Science, Okayama University, 3-1-1, Tsushimanaka, Kita-ku, Okayama 700-8530, Japan*

[4] *Laboratory of Ocean-Earth Life Evolution Research, Japan Agency for Marine-Earth Science and Technology, Kanagawa, Japan*

[5]*Department of Earth and Planetary Science, Tokyo Institute of Technology, Ookayama, Meguro-ku, Tokyo 152-8551, Japan.*

[6]*Department of Chemical Engineering and Materials Science, University of Minnesota, 421 Washington Ave SE, Minneapolis, MN 55455, USA*

[7]*Minnesota Supercomputing Institute, University of Minnesota, 117 Pleasant Street SE Minneapolis, MN 55455 USA*

# Abstract

Lakshtanov et al. (2007) showed that incorporation of aluminum and some water into $SiO_2$ significantly reduces the post-stishovite transition pressure in $SiO_2$. This discovery suggested that the ferroelastic post-stishovite transition in subducted MORB crust could be the source of reflectors/scatterers with low shear velocities observed in the mid to upper lower mantle. A few years later, a similar effect was observed in anhydrous Al-bearing silica. In this paper, we show by first principles static calculations and by molecular dynamics using inter-atomic potentials that hydrogen bonds and hydrogen mobility play a crucial role in lowering the post-stishovite transition pressure. A cooperative redistribution of hydrogen atoms is the main mechanism responsible for the transition pressure reduction in hydrous aluminous stishovite. The effect is enhanced by increasing hydrogen concentration. This perspective suggests a potential relationship between the depth of seismic scatterers and the water content in stishovite.

**Introduction**

Silica (SiO$_2$) is an abundant component of the Earth's crust but it is not expected to be present in normal peridotitic mantle. Nevertheless, one of its phase transitions has been implicated as cause of seismological reflectors/scatterers observed in the upper to middle layers of the lower mantle, ranging from 800 to 1800-km depth (e.g., Kawakatsu and Niu, 1994; Le Stunff et al., 1995; Kaneshima and Helffrich, 1999, 2003, 2009; Vinnik et al., 2001, Niu et al., 2003). Stishovite, the lowest pressure form of SiO$_2$ containing silicon in octahedral coordination, transforms to CaCl$_2$-type SiO$_2$ (post-stishovite transition) at ~50 GPa (~1,400 km depth) and room temperature [Kingma et al., 1995]. It is a second-order ferroelastic transition consisting of an orthorhombic distortion of tetragonal stishovite and displaying large shear modulus softening [Cohen, 1991; Karki et al., 1997; Carpenter et al., 2000; Tsuchiya et al., 2004]. The observed seismic heterogeneities may be reconciled with the post-stishovite phase transition in SiO$_2$ hosted in the subducted MORB crust (e.g., Kaneshima and Helffrich, 1999; Tsuchiya et al., 2004; Hirose et al., 2005) with ~25 mol% SiO$_2$. Recent high *P-T* experiments demonstrated that the post-stishovite transition occurs at about 70 GPa along the geotherm in pure SiO$_2$ [Nomura et al., 2010], accounting for the anomalies approximately at 1500 to 1800-km depth. Shallower seismic heterogeneities may be attributed to the transition in the Al-bearing SiO$_2$ phase, with or without water [Lakshtanov et al., 2007; Bolfan-Casanova et al., 2009]. Recent first principles calculations of acoustic velocities of stishovite and CaCl$_2$-type SiO$_2$ at lower mantle conditions showed surprising results, such as the $\Delta V_S/\Delta V_P \sim 2$ for this transition at mantle conditions [Yang and Wu, 2014].

Several other possibilities have been suggested to explain mid-lower mantle heterogeneities. The breakdown of phase D [Shieh et al., 1998], a dense hydrous magnesium silicate, into MgSiO$_3$ perovskite and water could cause hydration of mid-lower mantle around 1500-km depth [Ohtani et al., 2001b; Niu et al., 2003], but it cannot explain the shallower seismic heterogeneities. The tetragonal-cubic transition in CaSiO$_3$-rich perovskite has also been proposed to cause mid-lower mantle anomalies [Kurashina et al., 2004]. However, the phase boundary and the shear velocity discontinuity across the transition are still under debate [Li et al., 2006a, 2006b; Stixrude et al., 2007; Komabayashi et al., 2007]. Moreover, such phase transition in Ca-perovskite is expected to occur throughout the entire lower mantle considering the variation in temperature of subducted slabs, while seismic anomalies are found only above 1800-km depth. A spin crossover in $Fe^{3+}$ in Al-bearing phase D expected to be present in MORB has more recently been proposed also [Chang et al., 2013]. Iron spin crossovers in lower mantle minerals occur in broad pressure ranges [Badro et al., 2003, 2004; Tsuchiya et al., 2006; Lin et al., 2007; Wentzcovitch et al., 2009] and their effects on seismic velocities seem to reach a maximum between 1200 km and 1800 km depth [Wentzcovitch et al., 2009; Catalli et al., 2010; Hsu et al., 2011; Wu and Wentzcovitch, 2014], which correlates well with the depth range of observed scatters. However, spin crossovers affect most strongly bulk and longitudinal velocities [Marquardt et al., 2009; Wu et al., 2013]. Their effect on shear velocities is still debated [Crowhurst et al., 2008; Antonangeli et al., 2011].

In this paper we report a computational study of the post-stishovite transition in anhydrous and hydrous aluminous $SiO_2$. Our goal is to understand the roles that alumina and water play separately on the transition and implications of seismological observations in light of these results. First we investigate by first principles the nature of the aluminous defect, with and without hydrogen, in supercells containing up to 163 atoms. We uncover a cooperative effect associated with hydrogen bonds (H-bonds) and hydrogen motion. After characterizing these defects, we perform large-scale molecular dynamics simulations using inter-atomic potentials reproducing experimental conditions more closely.

**Results and Discussion**

*First principles defect calculations*

First, we address the combined effect of aluminum and hydrogen by means of $SiO_2 \leftrightarrow AlOOH$ substitutions. Pure AlOOH exists in the $CaCl_2$-type phase, the δ-AlOOH phase [Suzuki et al., 2000; Ohtani et al., 2001a], which is highly soluble in stishovite [Pawley et al, 1993; Chung and Kagi, 2002]. Calculations were performed in 2*2*2 (49 atoms) and 3*3*3 (163 atoms) supercells with one AlOOH unit/cell (6.25 mol% and 1.85 mol%, respectively) (See Appendix A for calculation details). Possible defect configurations are illustrated in Fig. 1. There are twelve hydrogen sites grouped in two symmetry related sets in the vicinity of an Al-octahedron: a fourfold degenerate site (H1) and an eightfold degenerate (H2). H1 and H2 hydrogens are bonded to apical and equatorial oxygens of the Al-octahedron, respectively. H2 hydrogens are more stable than H1s. Previously, H2 hydrogens were reported to be unstable in LDA calculations [Panero and Stixrude, 2004]. The enthalpy difference between these two types of hydrogens is –0.010 Ry at 0 GPa in the 2*2*2 supercell and decreases to -0.006 Ry at 50 GPa. Their relative populations are

$$\frac{N_2}{N_1} = \exp\left[-\left(\frac{(H_2 - H_1) - k_B T \ln 2}{k_B T}\right)\right],$$

where $H_1$ and $H_2$ are enthalpies of H1 and H2 hydrogen types. The ($k_B T \ln 2$) factor appears because H2 sites are twice as abundant as the H1 sites. For instance, at 1000 K, $N_1 \sim 9\%$ and $N_2 \sim 91\%$ at 0 GPa, and $N_1 \sim 17\%$ and $N_2 \sim 83\%$ at 50 GPa.

These calculations reveal a remarkably strong and concentration-dependent "chemical" effect of hydrogen on the post-stishovite transition (Fig. 2). Stishovite supercells with one AlOOH unit are orthorhombic even at 0 GPa. The distortion increases with pressure, and its magnitude depends on the hydrogen site/type and concentration, the distortion increasing with the latter. The strong distortion is caused by a hydrogen bond (H-bond) formed with the oxygen across the interstitial site. Depending on the hydrogen position, the H-bond aligns more along the **a** or the **b** axis, which determines whether the distortion produces **a>b** or **a<b** (Fig. 1(B)). The orthorhombic symmetry further splits H1 and H2 hydrogens into two sets each, twofold $H1_1$ and $H1_2$, and fourfold $H2_1$ and $H2_2$. $H1_1$ and $H2_1$ hydrogens make **a>b**, while $H1_2$ and $H2_2$ make **a<b**. This result is a consequence of the long range order and cooperative alignment of H-bonds in calculations containing a single AlOOH unit per supercell. Therefore, it is apparent that AlOOH-bearing stishovite should have hydrogens randomly distributed, mostly in H2 sites, in a way that stresses caused by one H-bond are opposed and canceled by stresses caused by all others, keeping the structure in the averaged undistorted tetragonal structure.

To illustrate this point, we prepared 4*2*2 supercells with two AlOOH units and two H-bonds. Two different configurations were generated: i) one $H2_1$ and one $H2_2$. This configuration keeps **a≈b** and is referred to as tetragonal. These hydrogens individually promote opposing distortions with cancelling effects. Beyond 50 GPa, this structure undergoes the usual orthorhombic distortion, similar to defect-free stishovite; and ii) two $H2_2$. This configuration produces **a<b**, the orthorhombic CaCl$_2$-type structure. Stresses caused by these H-bonds add coherently resulting in a large distortion even at 0 GPa. The orthorhombic structure is more stable at all pressures in these static calculations (see Appendix B). However, upon warming, configuration entropy contributes to the stabilization of the tetragonal structure, because of the higher number of hydrogen configurations in the tetragonal structure. Qualitatively, our approximate treatment of configuration entropy (see Appendix B) indicates that increasing temperature favors the tetragonal structure over the orthorhombic one producing a positive Clapeyron slope (See Fig. 7 in Appendix B) and *the post-stishovite transition in AlOOH-bearing SiO$_2$ is accompanied by a redistribution of hydrogens*. Therefore, the mechanism of the hydrous post-stishovite transition is different from that in pure stishovite; the post-stishovite transition induced by hydrogen redistribution is a first-order transition accompanied by an entropy discontinuity (see Appendix C).

We also investigated effects of incorporation of Al$_2$O$_3$ (see Appendix D). However, the post-stishovite transition is not affected by the presence of Al$_2$O$_3$. Our results reveal that the presence of hydrogen is crucial in shifting the transition pressure. This finding does not support conclusions from recent *in situ* x-ray diffraction experiments with a laser-heated diamond-anvil cell on the post-stishovite transition on the anhydrous Al-bearing phase [Bolfan-Casanova, 2009]. This experiment observed a substantial decrease in the transition pressure with the addition of 4 wt. % Al$_2$O$_3$. These authors state that their samples, which were prepared from a glass or a gel, "should contain a negligible amount of water". However, no evidence (e.g., infrared spectra or secondary ion mass spectrometry analyses) was provided to substantiate this statement. Therefore, the possibility remains that some amount of hydrogen was present in their samples, as required according to our calculations for the observed decrease in transition pressure.

*Molecular dynamics simulations*

These first-principles results indicate that, superposed to the ferroelastic post-stishovite transition displaying shear modulus softening, there is a strong chemical effect caused by H-bonds. The latter decreases the transition pressure by an amount that depends on hydrogen concentration and temperature (i.e., hydrogen configuration). Throughout the hydrous post-stishovite transition, the hydrogen configuration must change, i.e., hydrogen ions must be mobile to explore configuration space. If not, hydrogen might not affect the post-stishovite transition boundary.

To investigate hydrogen dynamics in the post-stishovite transition, we performed MD simulations in 8*8*12 supercells (4,704 atoms) (see Appendix E for details). The goal is to confirm the correlation between structural distortion and hydrogen motion. The initial configurations consisted of 96 uniformly distributed aluminum atoms with hydrogen atoms randomly distributed among the twelve sites surrounding each aluminum octahedron (6.25 mol% of AlOOH). The relaxed configurations are all tetragonal and metastable at low temperatures. The lowest enthalpy structure/configuration at 0 K, is orthorhombic at all pressures with equal hydrogen populations in $H2_1$ (or $H2_2$) sites.

Figure 3 shows hydrogen trajectories and the lattice constant ratio, b/a, in some of these MD runs. Below ~500 K, hydrogen motion is severely suppressed. The effect of AlOOH is negligible, and the post-stishovite transition pressure is ~50-60 GPa (Figs. 3(A) and 4), essentially the same as in pure silica (see Fig. 4). The suppression of hydrogen motion does not change the transition pressure or the nature of the post-stishovite transition. At these low temperatures, the transition is induced by the normal mechanical/dynamical instability observed in stishovite. As temperature increases, hydrogen mobility also increases. Short-range hopping, mostly between sites surrounding a single oxygen octahedron, is followed by the orthorhombic structural distortion at ~40-50 GPa. This is thermal "annealing" of the metastable structure, not a thermodynamic phase transition. Hydrogen ions were initially "trapped" in the tetragonal configuration and now find their way to more stable sites in the orthorhombic structure lowering the free energy. At ~900 K, this effect is strongest and this type of hydrogen hopping is more frequent. These orthorhombic distortions, although not severe, occur at pressures as low as ~20-30 GPa (Figs. 3(B) and 4). At 1100~1300 K, the structure remains tetragonal to higher pressures (Figs. 3(C) and 4). Hydrogen motion is significant, especially along the c axis. We see here the entropic stabilization of the tetragonal phase. This is the real pressure-induced hydrous post-stishovite transition. This is a first order phase transition involving redistribution of hydrogens. At 1500 K, the transition pressure is ~50-60 GPa (Figs. 3(D) and 4). At this temperature, there is no significant effect of hydrogens and the normal ferroeslastic post-stishovite transition takes place (Fig. 3(D)). Hydrogen motion no longer plays a role in assisting the phase transition that is now driven by the elastic instability. Hydrogen atoms necessarily redistribute with structural distortion and this transition is now accompanied by entropy and volume change as in a normal first order transition.

Although we only report results of few simulations, similar runs starting from distinct but equivalent configurations produced completely similar results (see right panels in Fig. 3). There are large uncertainties in transition pressures caused by the initial atomic configuration in the MD runs (see error bars in Fig. 4). Therefore, one must not take transition pressures and temperatures literally, especially at low temperatures. Besides, aluminum is uniformly distributed in the simulation cells instead of randomly distributed. Our goal here is to investigate the relationship between hydrogen distribution/motion and structural distortions. Nevertheless, some lessons learned apply to interpretation of experiments. For instance, tetragonal hydrous stishovite samples at low temperatures and pressures might be metastable. The tetragonal phase can exist metastably because of the first-order nature of the hydrous post-stishovite (see Appendix C). Upon initial slow moderate warming and cooling cycles, hydrous stishovite might be annealed into a thermodynamically stable orthorhombic phase at low temperatures. In hydrous aluminous stishovite, further aluminum disorder – not considered here – might help to stabilize the tetragonal phase. But clearly, if hydrogen motion is suppressed at low temperatures, AlOOH incorporation should not affect the transition pressure significantly. *Change in the post-stishovite transition pressure implies that hydrogen is mobile.*

At high temperatures where hydrogen ions are mobile, we can also anticipate that the post-stishovite transition in hydrous aluminous stishovite will have a larger Clapeyron slope than the transition in pure $SiO_2$ (Fig. 4), i.e., > 6 MPa/K [Tsuchiya et al., 2004]. In addition to the decrease in vibrational entropy across the regular post-stishovite transition in silica, there is additional reduction in hydrogen configuration entropy. Surely the Clapeyron slope of the hydrous post-stishovite transition should depend monotonically on the AlOOH concentration. Summarized results of MD simulations on a supercell containing 6.25 mol% AlOOH reported in Fig. 4 suggest a Clapeyron slope of ~ 50 ± 30 MPa/K. This is much steeper than that of AlOOH-free stishovite, because there is an additional effect of hydrogen configuration entropy (see Appendix B).

*Further insights from elasticity*

In pure $SiO_2$, the post-stishovite transition at 300 K results from mechanical and dynamical instabilities at ~50 GPa [Cohen, 1991; Karki et al., 1997; Tsuchiya et al., 2004]. A zone-center optical $B_{1g}$ mode, corresponding to an octahedral rotation, softens under pressure. This softening is accompanied by the breakdown of one of Born stability criteria, $c_{11}-c_{12}$. Fig. 5 shows acoustic velocities calculated using the tetragonal 4*2*2 supercells (6.25 mol% of AlOOH). Results are compared with velocities obtained by Brillouin scattering for samples containing 5 wt% of $Al_2O_3$ and possibly water [Lakshtanov et al., 2007]. The dip of acoustic velocity reported by experiments and by high temperature calculations [Yang and Wu, 2014] is well reproduced. Correspondingly, the calculated shear modulus has a dip across the transition. The underestimated velocity in these calculations originates primarily in the use of GGA, which overestimates volume and underestimates elastic properties. The use of GGA was preferred because of the presence of hydrogen bonds and their importance in this transition. The level of agreement with measured velocities is similar to other GGA calculations, despite the static nature of the hydrogen ions. Therefore, velocities of pure and hydrous aluminous stishovite do not appear dramatically different, but the hydrous supercell distorts more readily under pressure. This behavior suggests that in Brillouin experiments, hydrogen motion at room temperature was not vigorous, *except at the transition, where hydrogen configurations must change with change of symmetry if the transition pressure is affected*. Experimentally, the transformation occurs at 25 GPa, as seen in Fig. 5. At this point $c_{11}-c_{12}$, one of the Born criteria for structural stability, is still large (see Appendix C). Therefore, the post-stishovite transition in these samples is no longer a second-order ferroelastic transition but it is a regular first-order enthalpy driven (ΔV<0) transition as discussed previously, without elastic instability. First-order phase transitions can also produce shear wave anomalies, as documented well experimentally [Li and Weidner, 2008].

These calculations suggest that hydrogen might affect considerably the elastic properties of stishovite at temperatures where hydrogen is mobile, though this might not manifest in ultra-sound resonance or Brillouin scattering experiments. Small deformations of the tetragonal cell reduce degeneracies of hydrogen configurations, changing the equilibrium populations of hydrogen sites. If the time frame of the applied load is comparable to hydrogen relaxation times, rearrangement of hydrogen ions will take place, which will significantly affect elastic properties. Although seismic waves produce small stresses, their periods are large enough (> 1 sec) and should allow hydrogen relaxation into equilibrium configurations at mantle temperatures. Such anelastic deformation, termed the Snoek relaxation, is well-documented for bcc metals containing a small concentration of small, highly mobile interstitial solute atoms including H, C, N or O [Blanter et al., 2007]. One should therefore measure/calculate the relaxed elastic moduli, which goes beyond the scope of this work. These relaxed moduli should be considerably reduced.

*Summary*

We have shown that H-bond formation in hydrous aluminous stishovite can significantly reduce the transition pressure of the post-stishovite transition. Hydrogen bonds and hydrogen mobility affect the transition pressure because equilibrium populations of hydrogen sites differ in tetragonal and orthorhombic structures. If hydrogen motion is suppressed, incorporation of AlOOH should not affect the transition pressure significantly. This understanding also points to the motion of hydrogen ions as an important source of anelasticity and, hence, reduction, dispersion, and attenuation of seismic waves in hydrous aluminous stishovite. Although the real system is far more complex than those simulated and more complex atomistic phenomena cannot be ruled out, our results suggest that $Al_2O_3$ alone does not affect the transition pressure, while the effect of water is clearly identified, significant, and concentration dependent. Some experiments revealed a dependence of transition pressure on $Al_2O_3$ concentration. However, our results suggest that a trace of water might have accompanied $Al_2O_3$ in these experiments.

So far, several possible origins for seismic reflectors/scatterers observed in upper to middle lower mantle (800 to 1500-km depth) have been proposed: the breakdown of phase D, the tetragonal-cubic phase transition of Ca-perovskite, the post-stishovite transition in pure $SiO_2$, and the spin transition if $Fe^{3+}$ in phase D. Our results suggest that the hydrous post-stishovite transition is also a likely candidate. The depth/pressure range of the scatterers might be consequence of variable $H_2O$ in stishovite in MORB crusts subducted into the lower mantle.

While the role of stishovite in carrying water into the lower mantle has been stressed [Litasov and Ohtani, 2005], the amount of $H_2O$ incorporated in Al-bearing stishovite has remained controversial. The solubility of $Al_2O_3$ strongly depends on temperature and is possibly <1 wt% below 1400 K as indicated by [Ono et al., 2002], which limits the amount of $H_2O$ transported by stishovite [Lakshtanov et al., 2007; Ono et al., 2002]. Nevertheless, this study indicates that in hydrous stishovite a phase transition could occur in the upper to middle layers of the lower mantle and the transition pressure should depend on the $H_2O$ content. Shear velocity anomalies accompanying this transition could be enhanced by relaxation of hydrogen positions at high temperatures.

**Appendix A: First-principles calculations**

These density functional calculations were performed using the PBE-type generalized gradient approximation (GGA) [Perdew et al., 1996]. We have used Vanderbilt pseudopotentials [Vanderbilt, 1990] for all elements. We used the same cut-off radius for all angular momenta for each element: 1.6, 2.0,1.4, and 0.6 a.u for Si, Al, O, and H, respectively. The plane-wave energy cut-off was 40 Ry. k-point meshes were 4*4*6, 2*2*4, 1*2*4, 2*2*2 and 2*2*2 for 1*1*1 (i.e., pure stishovite), 2*2*2, 4*2*2, 2*2*4, and 3*3*3 supercells, respectively. Total energies converged to less than 1 mRy/atom. Structural optimizations at arbitrary target pressures were performed using variable cell shape molecular dynamics [Wentzcovitch, 1991]. To determine the phase boundary of the post-stishovite transition in pure $SiO_2$, quasi-harmonic approximation was applied, combined with phonon frequencies calculated by density-functional-perturbation theory. All first-principles calculations were performed by using the Quantum ESPRESSO software distribution [Giannozzi et al., 2009].

**Appendix B: Approximate treatment of effects of hydrogen configurations**

First-principles calculations of the effect of hydrogen configurations were conducted on 4*2*2 supercells with two AlOOH units (i.e. two H-bonds) in order to generate both tetragonal and orthorhombic structures. The supercells are tetragonal when two hydrogen ions are placed on $H2_1$ and $H2_2$ sites (see Fig. 1) or orthorhombic when they are both place on $H2_1$ or $H2_2$. Therefore, the number of hydrogen configurations in the tetragonal structure is larger than in the orthorhombic structure. Similarly, configuration entropy in the tetragonal structure is higher than in the orthorhombic structure. A quantitative estimate of the configuration entropy should start from the partition function sampling all possible hydrogen configurations for supercells which should be much larger than the supercells used in the present study. This is unfeasible by first principles. Instead, we qualitatively address configuration entropy in the relatively small 4*2*2 supercells in the following way: around each Al-octahedron, there are four $H2_1$ and $H2_2$ sites. In the tetragonal structure, two hydrogens are distributed randomly over these sites. There are eight possible sites for each hydrogen around each Al-octahedron, therefore, there are 64 H configurations for a cell with two Al ions. In contrast, in the orthorhombic structure two hydrogens are distributed randomly over $H2_1$ or $H2_2$ sites only (leading to a>b or a<b), resulting in 4*4, i.e., 16 configurations. In this way, configuration entropy is estimated to be approximately $k_B \ln 64$ and $k_B \ln 16$ for tetragonal and orthorhombic structures, respectively ($k_B$ is the Boltzmann constant). With these entropies, we can estimate Gibbs free energies for tetragonal and orthorhombic phases at finite temperatures: $G_{tetra(ortho)}(P,T) = H_{tetra(ortho)}(P) - TS_{tetra(ortho)}$, where H is enthalpy for each structure. At 0 K, Gibbs free energy is equivalent to enthalpy. Differences in Gibbs free energies between tetragonal and orthorhombic structures at several temperatures ($\Delta G = G_{ortho} - G_{tetra}$) are shown in Fig. 7. Below approximately 300 K, $\Delta G$ is always negative and the orthorhombic phase is stable at all pressures. Above ~300 K $\Delta G$ is positive below a critical pressure and the post-stishovite transformation takes place at high pressures. Transition pressures are ~ 6, 24, 32, and 38 GPa at 500, 1000, 1500, and 2000 K, respectively. The post-stishovite transition has positive Clapeyron slope in the absence hydrogen or of hydrogen redistribution. This positive slope is further enhanced because of the extra reduction of entropy caused by hydrogen redistribution.

**Appendix C: First-order versus second-order transition**

The post-stishovite transition in pure $SiO_2$ has been well investigated [Cohen, 1991; Karki et al., 1997; Carpenter et al., 2000; Tsuchiya et al., 2004; Yang and Wu, 2014]. Beyond the transition pressure, the zone-center soft mode induces tetrahedral rotations. Correspondingly, one of the Born stability criteria breaks down ($c_{11}-c_{12}<0$), leading to a mechanical instability. This instability drives the tetragonal to orthorhombic transition. Across this transition, lattice constants change continuously and there is (almost) no volume or entropy discontinuities. This is a second order ferroelastic transition. Also, in Al-bearing $SiO_2$, i.e., without reordering of oxygen vacancies and without hydrogen ions, the post-stishovite transition occurs when the Born stability criteria breaks down (Figs. 6 and 8); this is also a second-order transition.

However, the presence of mobile hydrogen changes the nature of this transition. Below ~500 K the free energy of the orthorhombic phase is always lower than that of the tetragonal phase (Fig. 7). Above ~500 K, hydrogen configuration entropy contributes further to the stabilization of the tetragonal phase with respect to the orthorhombic phase. Under pressure, redistribution of hydrogens further helps to induce the post-stishovite transition. Now lattice constants a and b change discontinuously (Fig. 9) at the transition pressure that is lower than that of the ferroelastic transition in pure $SiO_2$. But the Born stability criteria, $c_{11}$-$c_{12}$, of the tetragonal phase does not break down until ~50 GPa (Fig. 8). The presence of water hardly affects this criterion. All these indicators point to a first-order post-stishovite transition in the presence of water, in contrast to the second-order transition in anhydrous stishovite.

Because of the first-order transition, hysteresis is possible experimentally and in MD simulations, and the tetragonal phase may remain metastable in the stability field of the orthorhombic phase at low temperatures where hydrogen motion is severely suppressed. This is what is happening around and below ~ 500 K in Fig. 4. The post-stishovite transition under compression occurs not by hydrogen redistribution at this temperature but by the breakdown of a Born stability criterion, $c_{11}$-$c_{12}$ (Fig. 8), at ~50 GPa. If the transformation occurs without hydrogen rearrangement, the transition is still a second order ferroelastic transition.

This insight into the nature of the hydrous post-stishovite transition might be relevant for experimental investigations. Experiments in Lakshtanov et al. (2007) indicated that the sample at ambient condition was tetragonal. This might seem inconsistent with our qualitative results (see Fig. 7 suggesting that the tetragonal phase might be metastable at ambient condition). However, the experimental sample was prepared at 20 GPa and 1800 C. At these conditions the sample should be tetragonal (see Figs. 4 and 7). After quenching to ambient conditions the tetragonal sample might have been kept metastable because of structural hysteresis. Unless hydrogen is mobile, the required rearrangement to induce the hydrous post-stishovite transition does not take place and the transition remains ferroelastic in nature. In the same work the sample was *annealed* beyond 20 GPa to relax deviatoric stresses in the sample. Our study suggests that if the sample had not been annealed, the second-order transition might have been observed at ~50 GPa. In summary, annealing should be critical in the determination of the hydrous post-stishovite transition boundary.

**Appendix D: Effect of $Al_2O_3$ or oxygen vacancies**

To investigate the effect of $Al_2O_3$, we replaced $2SiO_2$ with $Al_2O_3$. This defect consists of two $Al_{Si}'$ and one $V_O^{\bullet\bullet}$ (assigned charges are only formal), i.e., the anhydrous defect. We consider two defect configurations: i) two corner-sharing and ii) two edge-sharing Al-octahedra with an oxygen vacancy in between (see Fig. 6). We carried out static calculations for both types in 4*2*2 and 2*2*4 supercells with 95 atoms. The enthalpy of the corner-sharing defects is essentially independent of the supercell used, while the edge-sharing defects form a chain in the 4*2*2 supercell, which makes this configuration less stable than in the 2*2*4 supercell. Results using the latter should be more accurate.

At 0 GPa, the edge-sharing defect has lower enthalpy, but pressure stabilizes the corner sharing defect beyond 40 GPa. Supercells with both types of defects display an orthorhombic distortion beyond 50 GPa, similar to aluminum-free stishovite (Fig. 6). Therefore, the post-stishovite transition pressure does not appear to be affected by the presence of these defects. This conclusion is consistent with that of a previous calculation [Panero, 2006]. Supercells with isolated alumimum octahedra should be less distorted than those investigated here, affecting less the post-stishovite transition.

**Appendix E: Molecular dynamics (MD) simulations**

N-P-T ensemble MD simulations were conducted at temperatures of 300, 500, 700, 900, 1100, 1300, and 1500 K at 0, 10, 20, 30, 40, 50, and 60 GPa using 8*8*12 supercells (4704 atoms) containing 6.25% AlOOH. Two-body interatomic potentials were used. Functional forms and parameters for these two-body interatomic potentials have been reported in Nakano et al. (2003). In the initial geometry, 96 aluminum atoms were uniformly distributed. 96 hydrogens were distributed randomly around aluminum octahedra so that initial phases were tetragonal. Variable cell shape simulations were allowed to evolve and equilibrate for 0.2 nanoseconds to achieve thermodynamic equilibrium. Hydrogen motion was monitored for 0.01 nanosenconds after cell equilibrations.

**Acknowledgments**

The authors thank David Kohlstedt, Justin Revenaugh, and George Helffrich for useful discussions. This work was supported by NSF under grants EAR-1161023 and EAR-1348066. Computations were performed at the Minnesota Supercomputing Institute (MSI) and in the Blue Waters system at NCSA.

# References

Antonangeli, D., Siebert, J., Aracne, C. M., Farber, D., Bosak, A., Hoesch, M., Krisch, M., Ryerson, F. J., Figuet, G., and Badro, J., (2011) Spin Crossover in Ferropericlase at High Pressure: A Seismologically Transparent Transition?, *Science* 331, 64-67.

Badro, J., Fiquet, G., Guyot, F., Rueff, J. P., Struzhkin, V. V., Vanko, G., and Monaco, G. (2003) Iron partitioning in Earth's mantle: Toward a deep lower mantle discontinuity, *Science* 300, 789-791.

Badro, J., Rueff, J. P., Vanko, G., Monaco, G., Fiquet, G., and Guyot, F. (2004) Electronic transitions in perovskite: Possible nonconvecting layers in the lower mantle, *Science* 305, 383-386.

Blanter, M. S., Golovin, I. S., Neuhäuser, H., Sinning, H. R. (2007) *Internal Friction in Metallic Materials: A Handbook*, Springer Series in Materials Science, Springer, New York, pp. 12-27.

Bolfan-Casanova, N., Andrault, D., Amiguet, E., Guignot, N. (2009) Equation of state and post-stishovite transformation of Al-bearing silica up to 100 GPa and 3000 K. *Phys. Earth Planet. Int.* 174, 70-77.

Carpenter, M. A., Hemley, R. J., Mao, H. K. (2000) High-pressure elasticity of stishovite and the $P4_2/mnm \leftrightarrow Pnnm$ phase transition. *J. Geophys. Res.* 105, 10807-10816.

Catalli, K., Shim, S. H., Prakapenka, V. B., Zhao, J., Sturhahn, W., Chow, P., Xiao, Y., Liu, H., Cynn, H., and Evans, W. J. (2010) Spin state of ferric iron in $MgSiO_3$ perovskite and its effect on elastic properties, Earth Planet. Sci. Lett. 289, 68-75.

Chang, Y. Y., Jacobsen, S. D., Lin, J. F., Bina, C. R., Thomas, S. M., et al. (2013) *Earth Planet. Sci. Lett.* 382, 1-9.

Chung, J. I., Kagi, H. (2002) High concentration of water in stishovite in the MORB system. Geophys. Res. Lett. 29, 16.

Cohen, R. E. (1991) Bonding and elasticity of stishovite $SiO_2$ at high pressure: Linearized augmented plane wave calculations. *Am. Mineral.* 76, 733-742.

Crowhurst, J. C., Brown, J. M., Goncharov, A. F., and Jacobsen, S. D. (2008) Elasticity of (Mg,Fe)O through the spin transition of iron in the lower mantle, *Science* 319, 451-453.

Giannozzi, P. *et al.* (2009) QUANTUM ESPRESSO: a modular and open-source software project for quantum simulations of materials. *J. Phys. Condens. Matter.* 21, 395502.

Hirose, K., Takafuji, N., Sata, N., Ohishi, Y. (2005) Phase transition and density of subducted MORB crust in the lower mantle. *Earth Planet. Sci. Lett.* 237, 239-251.

Hsu, H., Blaha, P., Cococcioni, M., de Gironcoli, S., and Wentzcovitch, R. M. (2011) Spin-state crossover and hyperfine interactions of ferric iron in MgSiO3 perovskite, *Phys. Rev. Lett.* 106, 118501.

Kaneshima, S., Helffrich, G. (1999) Dipping Low-Velocity Layer in the Mid-Lower Mantle: Evidence for Geochemical Heterogeneity. *Science* 283, 1888-1891.

Kaneshima, S., Helffrich, G. (2003) Subparallel dipping heterogeneities in the mid-lower mantle. *J. Geophys Res* 108 (B5), 2272.

Kaneshima, S., Helffrich, G. (2009) Lower mantle scattering profiles and fabric below Pacific subduction zones. *Earth Planet. Sci. Lett.* 282, 234-239.

Karki, B. B., Warren, M. C., Stixrude, L., Ackland, G. J., Crain, J. (1997) Ab initio studies of high-pressure structural transformations in silica. *Phys. Rev. B* 55, 3465-3471.

Kawakatsu, H., Niu, F. L. (1994) Seimsic evidence for a 920-km discontinuity in the mantle. *Nature* 371, 301-305.

Kingma, K. J., Cohen, R. E., Hemley, R. J., Mao, H. K. (1995) Transformation of stishovite to a denser phase at lower-mantle pressures. *Nature* 374, 243-245.

Komabayashi, T., Hirose, K., Sata, N., Ohishi, Y., Dubrovinsky, L. S. (2007) Phase transition in $CaSiO_3$ perovskite. *Earth Planet. Sci. Lett.* 260, 564-549.

Kurashina, T., Hirose, K., Ono, S., Sata, N., Ohishi, Y. (2004) Phase transition in Al-bearing CaSiO$_3$ perovskite: implications for seismic discontinuities in the lower mantle, Phys. Earth Planet. Inter. 145, 67-74.

Lakshtanov, D. L., Sinogeikin, S. V., Litasov, K. D., Prakapenka, V. B., Hellwig, H., Wang, J., Sanches-Valle, C., Perrillat, J. P., Chen, B., Somayazulu, M., Li, J., Ohtani, E., Bass, J. D. (2007) The post-stishovite phase transition in hydrous alumina-bearing SiO$_2$ in the lower mantle of the earth. *Proc. Nat. Acad. Sci.* 104, 13588-13590.

Le Stunff, Y., Wicks, Jr. C. W., Romanowicz, B. (1995) P'P' Precursors Under Africa: Evidence for Mid-Mantle Reflectors. *Science* 270, 74-77.

Litasov, K. D., Ohtani, E. (2005) Phase relations in hydrous MORB at 18-28 GPa: implications for heterogeneity of the lower mantle. *Phys. Earth Planet. Inter.* 150, 239-263.

Marquardt, H., Speziale, S., Reuchmann, H. J., Frost, D. J., Schilling, F. R., and Garnero, E. J. (2009) Elastic Shear Anisotropy of Ferropericlase in Earth's Lower Mantle, *Science* 324, 224-226.

Niu. F., Kawakatsu. H., Fukao. Y. (2003) Seismic evidence for a chemical heterogeneity in the midmantle: A strong and slightly dipping seismic reflector beneath the Mariana subduction zone. *J Geophys Res* 108(B9), 2419.

Li, L., Weidner, D., Brodholt, J., Alfve, D., Price, G. D., Caracas, R., and Wentzcovitch, R. M. (2006a) Elasticity of CaSiO$_3$ perovskite at high pressure and high temperature. *Phys. Earth. Planet. Inter.* 155, 249–259.

Li, L., Weidner, D., Brodholt, J., Alfve, D., Price, G. D., Caracas, R., and Wentzcovitch, R. M. (2006b) Phase stability of CaSiO$_3$ perovskite at high pressure and temperature: Insights from ab initio molecular dynamics. *Phys. Earth Planet. Inter.* 155, 260-268.

Li, L., Weidner, D. J. (2008) Effect of phase transitions on compressional-wave velocities in the Earth's mantle. Nature 454, 984-986.

Lin, J. F., Vanko, G., Jacobsen, S. D., Iota, V., Struzhkin, V. V., Prakapenka, V. B., Kuznetsov, A., and Yoo C. S. (2007) Spin Transition Zone in Earth's Lower Mantle, Science 317, 1740-1743.

Nakano, M., Kawamura, K., Ichikawa, Y. (2003) Local structural information of Cs in smectite hydrates by means of an EXAFS study and molecular dynamics simulations. *Appl. Clay Sci.* 23, 15-23.

Nomura, R., Hirose, K., Sata, N., Ohishi, Y. (2010) Precise determination of post-stishovite phase transition boundary implications for seismic heterogeneities in the mid-lower mantle. *Phys. Earth Planet. Int.* 183, 104-109.

Ohtani, E., Litasov, K., Suzuki, A., Kondo, T. (2001a) Stability field of new hydrous phase, δ-AlOOH, with implications for water transport into the deep mantle, Geophys. Res. Lett. 28, 3991-3993.

Ohtani, E., Toma, M., Litasov, K,. Kudo, T., Suzuki, A. (2001b) Stability of dense hydrous magnesium silicate phases and water storage capacity in the transition zone and lower mantle. Phys. Earth Planet. Inter. 124, 105-117.

Ono, S., Hirose, K., Murakami, M., Isshiki, M. (2002) Post-stishovite phase boundary in $SiO_2$ determined by in situ X-ray observations. *Earth Planet. Sci. Lett.* 197, 187–192.

Panero, W. R., Stixrude, L. P. (2004) Hydrogen incorporation in stishovite at high pressure and symmetric hydrogen bonding in d-AlOOH. *Earth Planet. Sci. Lett.* 221, 421-431.

Panero, W. R. (2006) Aluminum incorporation in stishovite. *Geophys. Res. Lett.* 33, L20317.

Pawley, A. R., McMillan, P. F., Holloway, J. R. (1993) Hydrogen in Stishovite, with implications for Mantle Water Content. Science 261, 1024-1026.

Perdew, J. P., Burke, K., Ernzerhof, M. (1996) Generalized Gradient Approximation Made Simple. *Phys. Rev. Lett.* 77, 3865-3868.

Sano-Furukawa, A., Komatsu, K., Vanpeteghem, C. B., and Ohtani, E. (2008) Neutron diffraction study of δ-AlOOD at high pressure and its implication for symmetrization of the hydrogen bond. Am. Mineral. 93, 1558-1567.

Sano-Furukawa, A., Kagi, H., Nagai, T., Nakano, S., Fukura, S., Ushijima, D., Iizuka, R., Ohtani, E., and Yagi, T. (2009) Change in compressibility of δ-AlOOH and δ-AlOOD at high pressure: A study of isotope effect and hydrogen-bond symmetrization. Am. Mineral. 94, 1255-1261.

Shieh, S. R., Mao, H., Hemley, R. J., Ming, L. C. (1998) Decomposition of phase D in the lower mantle and the fate of dense hydrous silicates in subducting slabs. *Earth Planet. Sci. Lett.* 159, 13-23.

Stixrude, L., Lithgow-Bertelloni, C., Kiefer, B., Fumagalli, P. (2007) Phase stability and shear softening in $CaSiO_3$ perovskite at high pressure. *Phys. Rev. B* 75, 024108.

Suzuki, A., Ohtani, E., Kamada, T. (2000) A new hydrous phase d-AlOOH synthesized at 21 GPa and 1000 °C. Phys. Chem. Minerals 27, 689-693.

Tsuchiya, T., Caracas, R., Tsuchiya, J. (2004) First principles determination of the phase boundaries of high-pressure polymorphs of silica. *Geophys. Res. Lett.* 31, L11610.

Tsuchiya, T., Wentzcovitch, R. W., da Silva, C. R. S., and de Gironcoli, S. (2006) Spin transition in magnesiowustite in the Earth's lower mantle, *Phys Rev Lett* 96, 198501.

Tsuchiya, J., Tsuchiya, T. (2009) Elastic properties of δ-AlOOH under pressure: First-principles investigation. Phys. Earth Planet. Inter. 174, 122-127.

Vanderbilt, D. (1990) Soft self-consistent pseudopotentials in a generalized eigenvalue formalism. *Phys. Rev. B* 41, 7892-7895.

Vinnik, L., Kato, M., Kawakatsu, H. (2001) Search for seismic discontinuities in the lower mantle. *Geophys. J. Int.* 147, 41-56.

Wentzcovitch, R. M. (1991) Invariant molecular-dynamics approach to structural phase transitions. *Phys. Rev. B* 44, 2358-2361.

Wentzcovitch, R. W., Justo, J. F., Wu, Z., da Silva, C. R. S., Yuen, D. A., and Kohlstedt, D. (2009) Anomolous compressibility of ferropericlase throughout the iron spin crossover, Proc. Nat. Acad. Sci. 106, 8447–8452.

Wu, Z., Justo, F. J., and Wentzcovitch, R. W. (2013) Elastic anomalies in a spin crossover system: ferropericlase at lower mantle conditions, *Phys. Rev. Lett.* 110, 228501.

Wu, Z. and Wentzcovitch, R. W. (2014) Spin crossover in ferropericlase and velocity heterogeneities in Earth's lower mantle, Proc. Nat. Acad. Sci. 106, 10468–10472.

Yang, R., Wu, Z. (2014) Elastic properties of stishovite and the $CaCl_2$-type silica at the mantle temperature and pressure: an ab initio investigation, *Earth Planet. Sci. Lett.* 404, 14-21.

# Figure Legends

**Fig. 1:** (A) Possible equilibrium sites of hydrogen in aluminous stishovite. White spheres represent possible hydrogen sites. (B) Resulting orthorhombic distortions depending on hydrogen bond orientations.

**Fig. 2:** Calculated (GGA) pressure dependence of the lattice constants ratio, **b/a,** for 2*2*2 (6.25% AlOOH) and 3*3*3 (1.85% AlOOH) supercells of $SiO_2$ with one AlOOH. Inset shows hydrogen positions.

**Fig. 3:** (Left and middle panels) Hydrogen trajectories (represented by black lines) in MD simulations at 60 GPa viewed along [001] (left) and [010] (middle). Small light blue and purple spheres denote silicon and aluminum respectively. Large yellow spheres denote oxygens. (Right panels) Pressure dependence of **b/a** for three MD simulations containing 4,704 atoms starting from three different configurations containing 96 aluminum atoms uniformly distributed and hydrogens in different randomly chosen configurations.

**Fig. 4:** Phase diagram of $SiO_2$ with 6.25 mol% AlOOH obtained from the MD simulations. Red symbols denote transition pressures determined from the MD simulations illustrated in Fig. 3. Black symbols represent phase boundary of pure silica obtained by similar MD simulations. Black line shows results of first-principles quasiharmonic GGA calculations. The agreement of black symbols and black line demonstrates the predictive quality of the inter-atomic potentials. No transition pressure shift is seen in the simulations at low temperatures. By contrast, large pressure shifts are observed above 900 K. The presumed thermodynamic phase boundary is sketched by the black dashed line. Color ranges denote different styles of hydrogen motion: A) Orange represents the region where quantum motion should be important and classical results are only qualitatively valid. Therefore, symbols are semi-transparent. In this temperature range, hydrogen motion is highly suppressed, the transition displays ferroelastic character, and the transition pressure is essentially the same as in pure silica. B) Yellow corresponds to the region where hydrogen is somewhat mobile. The pressure shift of the post-stishovite transition, in general, depends on the simulation time and annealing schedule. C) Green and blue show temperature ranges where hydrogen motion is significant and change in hydrogen configuration

entropy affects the post-stishovite transition pressure. In this temperature range MD simulations give approximately the equilibrium thermodynamics phase boundary of the hydrous post-stishovite transition (dashed black/red lines). D) In the blue region hydrogen diffuses throughout the crystal and the material has super-ionic character. The red dashed line represents the simulation phase boundary displaying hysteresis. Black dashed line is the best estimate of the thermodynamics phase boundary. The difference between black and red dashed lines indicates hysteresis is significant in the simulations.

**Fig. 5:** Acoustic velocities along [110] obtained by static first principles calculations for the 4*2*2 supercell. $\rho V^2_{[110]} = (c_{11}-c_{12})$ for the tetragonal phase and $\rho V^2_{[110]} = 1/2 * ((c_{11}+c_{22}+2c_{66}) - ((c_{11}-c_{22})^2+4(c_{12}+c_{66})^2)^{1/2})$ for the orthorhombic phase. At 20 GPa, where the transition occurs, the elastic anomaly ($c_{11}-c_{22}$) is no longer present and the transition is no longer ferroelastic, but a regular enthalpy driven first order transition. Inset: Calculated shear modulus by Voigt-Reuss-Hill average.

**Figure 6:** Calculated pressure dependence of the lattice constants ratio, **b/a,** for the 2*2*4 supercell of aluminous stishovite (3.125% AlOOH). Dark and light blue polyhedra in the inset panels denote Si- and Al-polyhedra. Small red spheres represent oxygens.

**Figure 7:** First-principles Gibbs free energy difference between two configurations (4*2*2 supercells with 98 atoms) containing two hydrogen bonds per supercell. The tetragonal phase contains one $H2_1$ and one $H2_2$ hydrogens. The orthorhombic phase contains two $H2_2$ hydrogens.

**Figure 8:** Calculated Born stability criterion, $c_{11}-c_{12}$, for pure $SiO_2$, $SiO_2$ with 6.25% AlOOH, and $SiO_2$ with $Al_2O_3$.

**Figure 9**: Calculated lattice constants, a and b, for tetragonal and orthorhombic structures for 4*2*2 supercells with 98 atoms containing two hydrogen bonds per supercell as indicated in Fig. 7 caption.

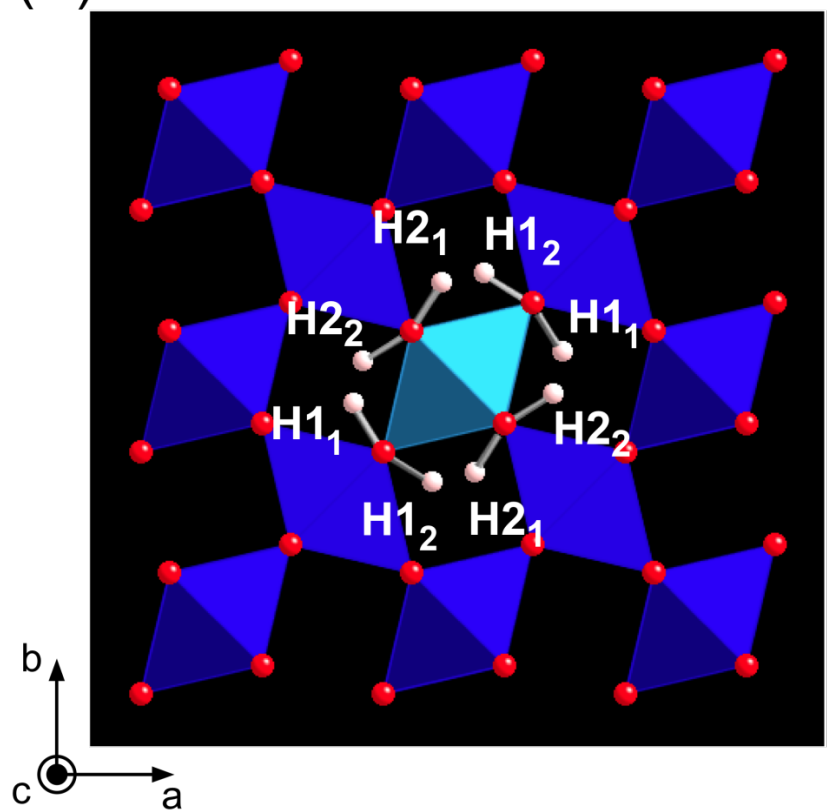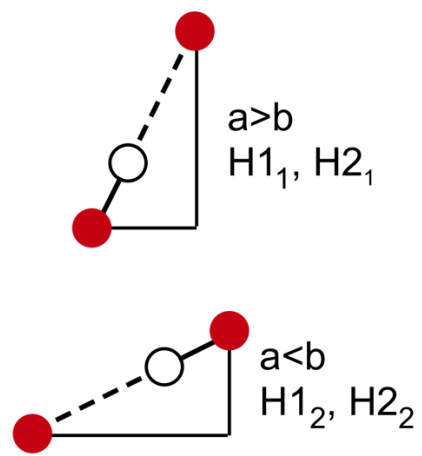

Fig. 1

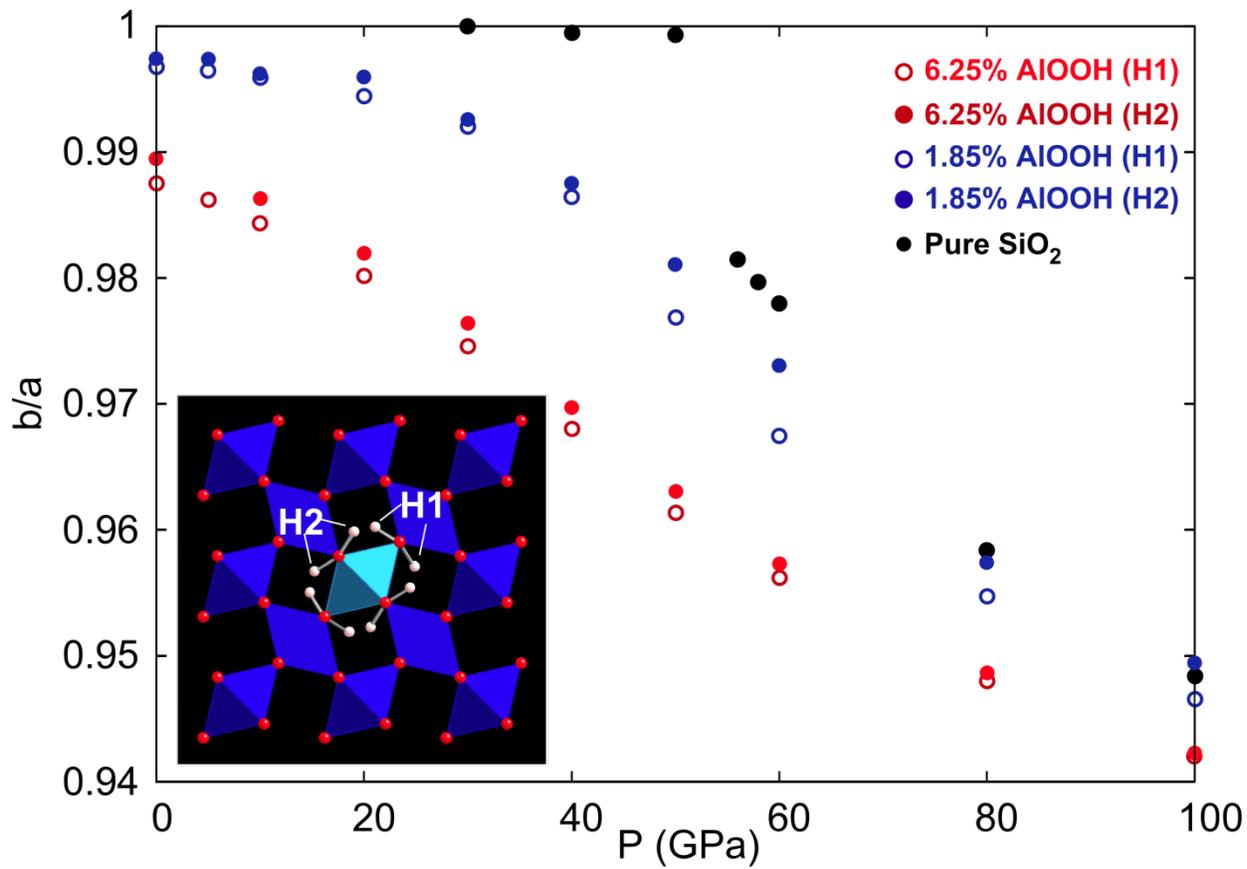

**Fig. 2**

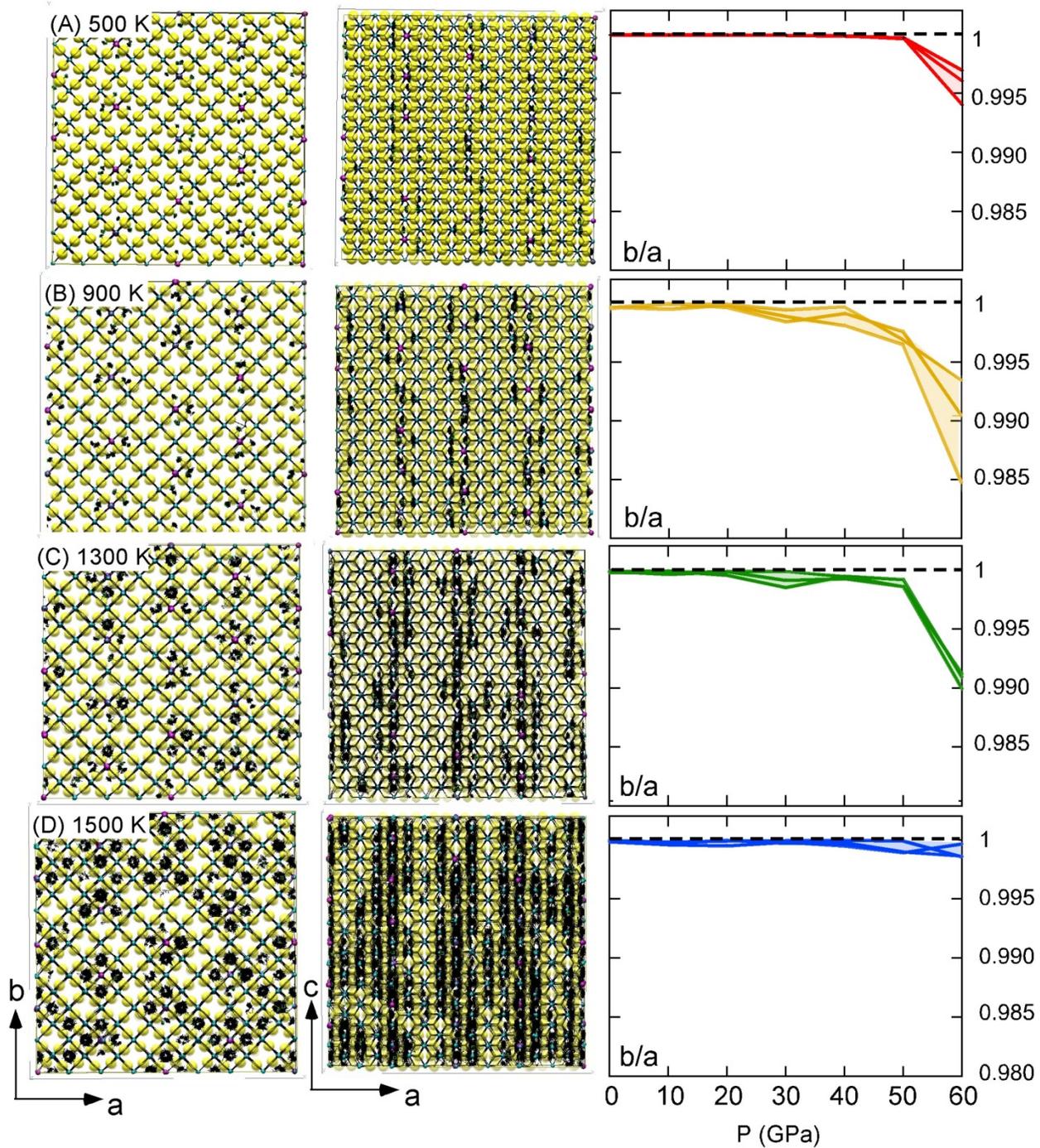

**Fig. 3**

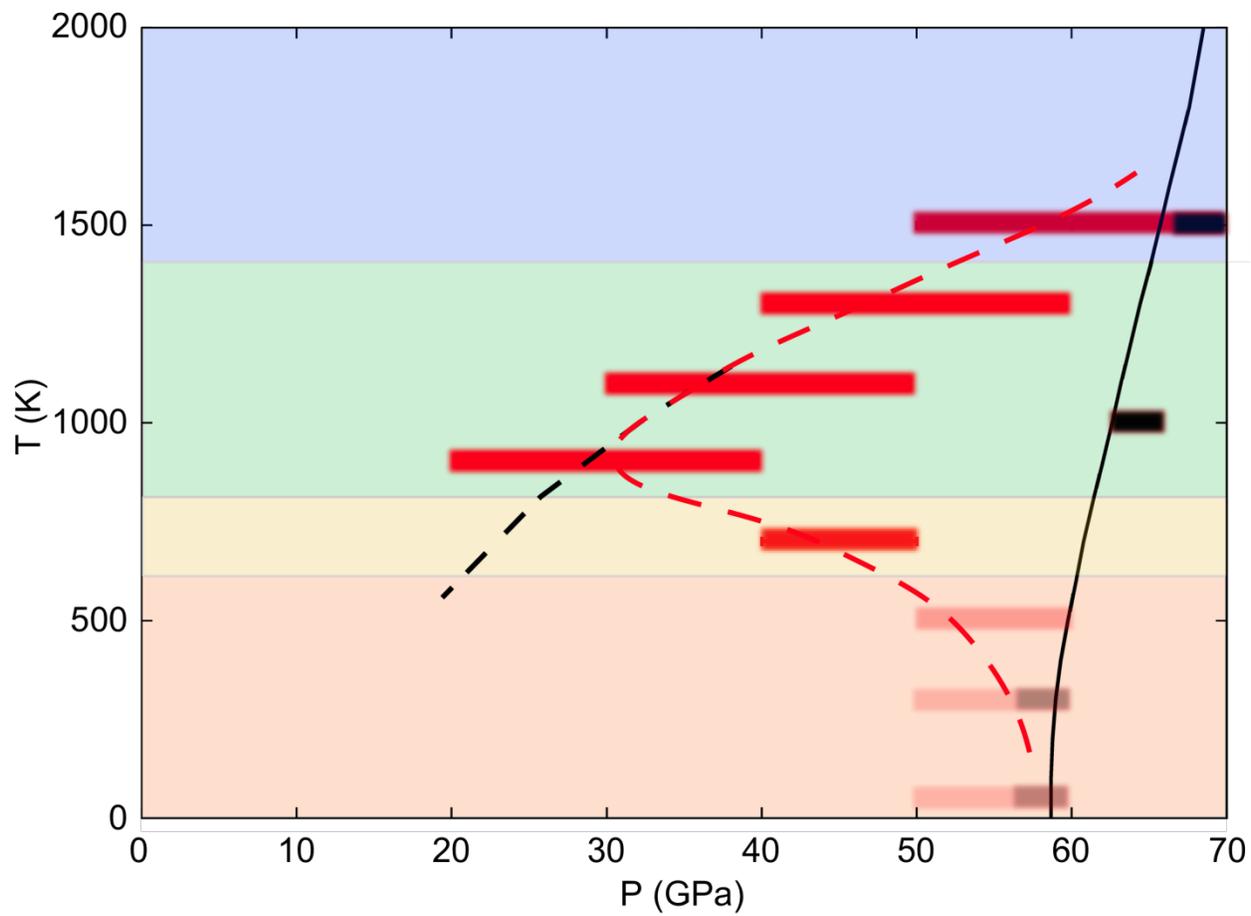

**Fig. 4**

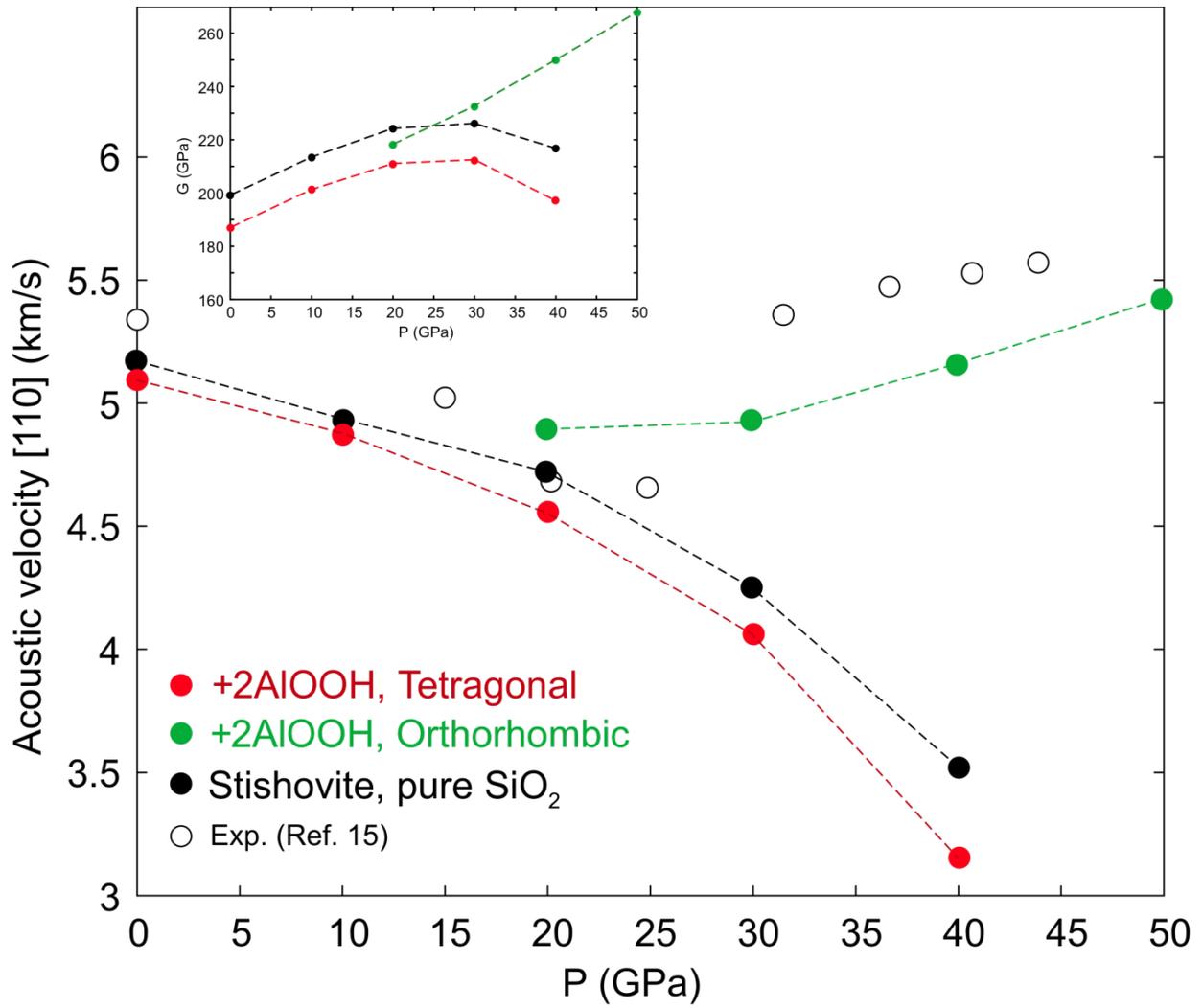

**Fig. 5**

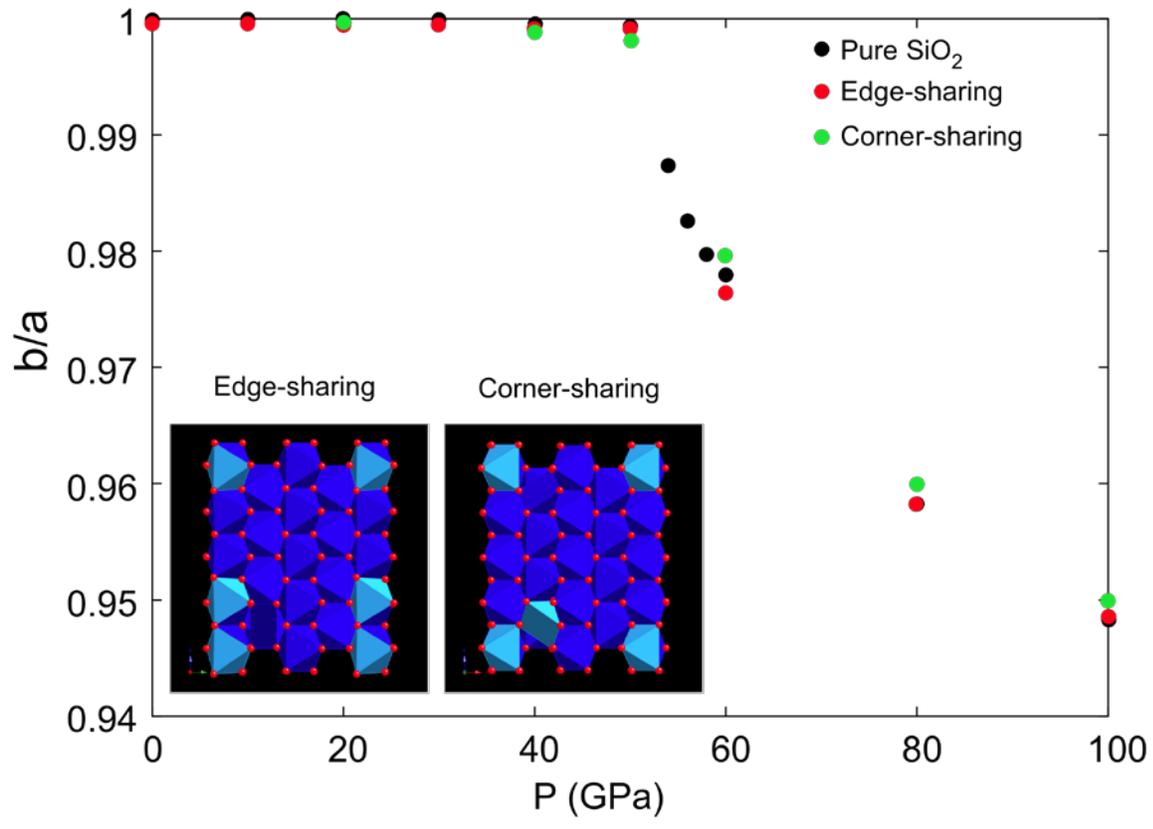

**Fig. 6**

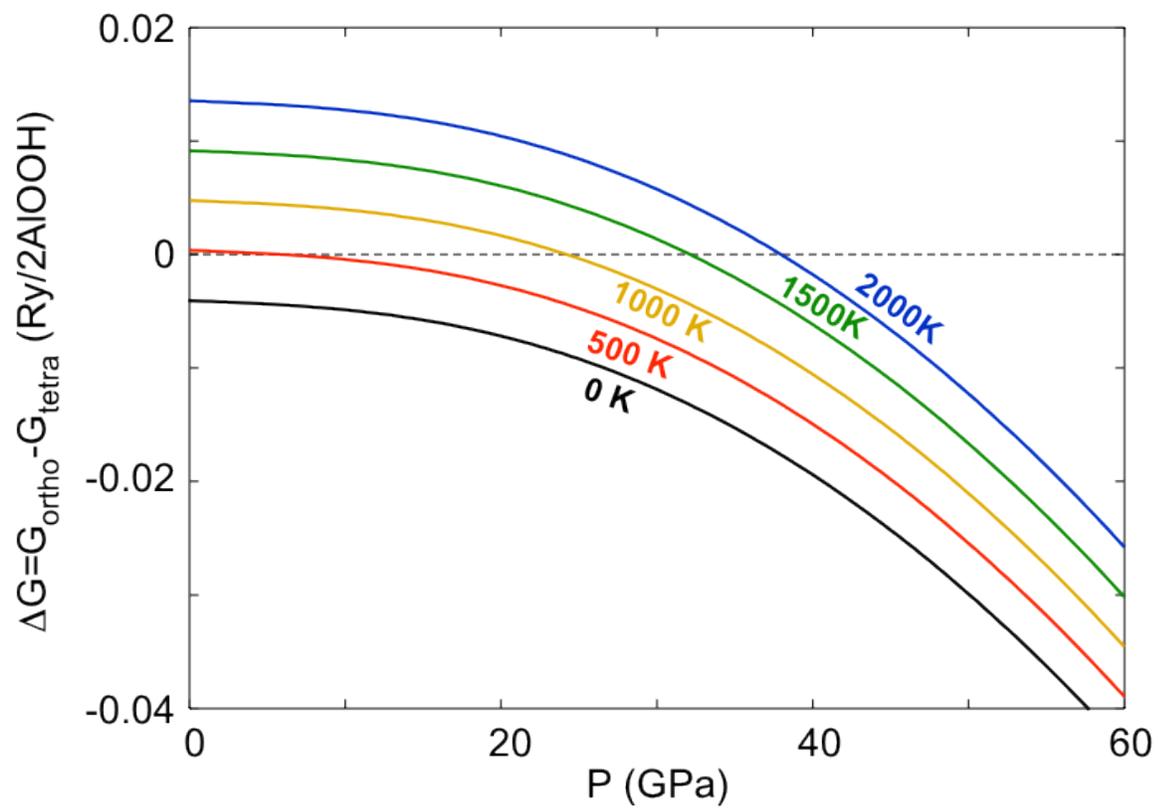

**Fig. 7**

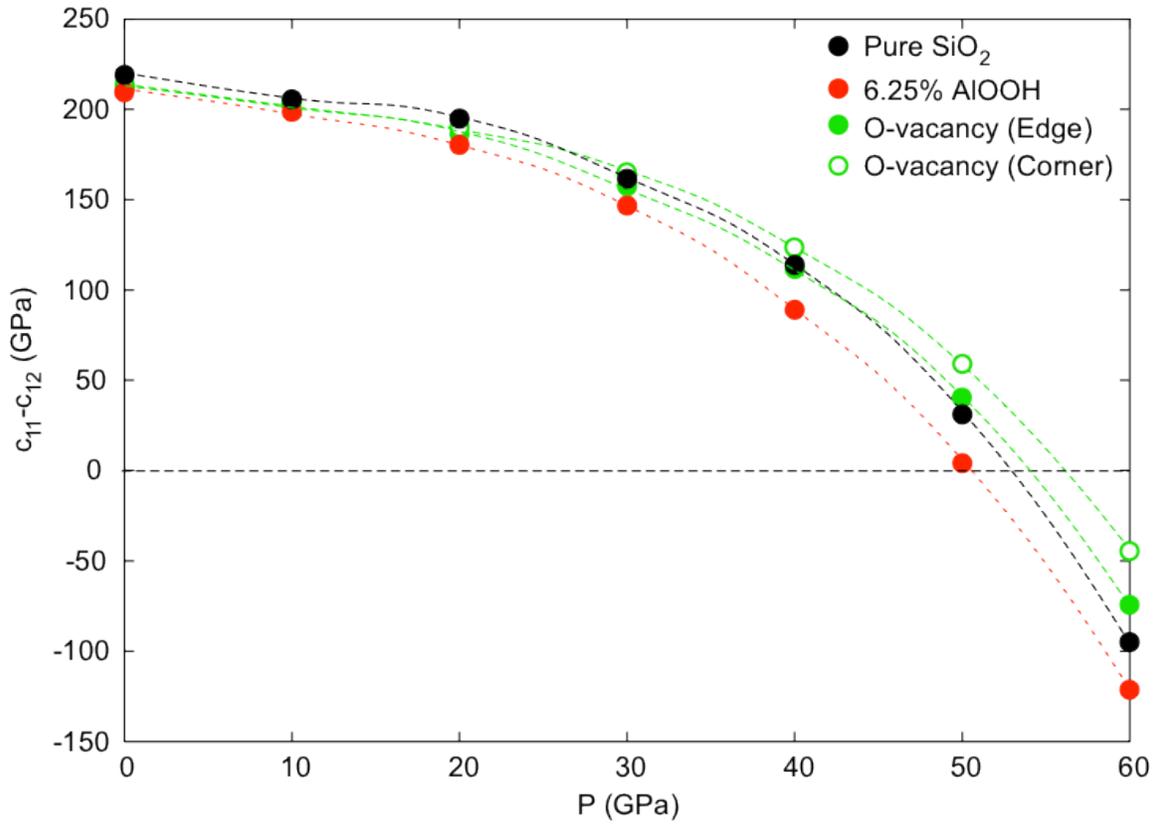

**Fig. 8**

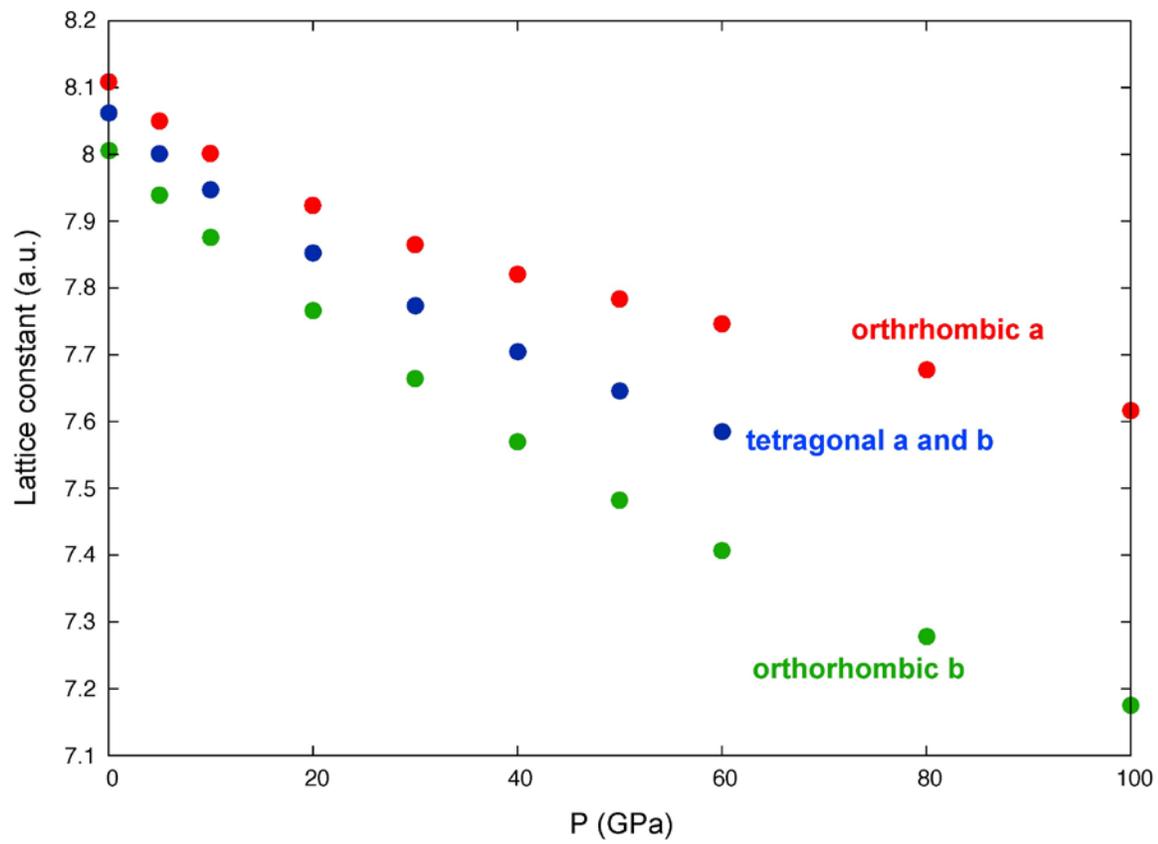

**Fig. 9**